\documentclass[prc,twocolumn,showpacs]{revtex4}

\date{March 28, 2002}
\usepackage{hyperref}
\usepackage{graphicx}
\usepackage{amssymb}
\usepackage{amsmath}

\newcommand{\xenon}[1][129]{{}^{#1}\mathrm{Xe}}
\newcommand{\tin}[1][]{{}^{#1}\mathrm{Sn}}

\begin{document}

\title{Compatibility of localized wave packets and unrestricted single
particle dynamics for cluster formation in nuclear collisions}

\author{Akira Ono}
\affiliation{Department of Physics, Tohoku University, Sendai
980-8578, Japan}
\author{S. Hudan}
\author{A. Chbihi}
\author{J. D. Frankland}
\affiliation{GANIL, 14076 Caen Cedex, France}

\begin{abstract}
Antisymmetrized molecular dynamics with quantum branching is
generalized so as to allow finite time duration of the unrestricted
coherent mean field propagation which is followed by the decoherence
into wave packets.  In this new model, the wave packet shrinking by
the mean field propagation is respected as well as the diffusion, so
that it predicts a one-body dynamics similar to that in mean field
models.  The shrinking effect is expected to change the diffusion
property of nucleons in nuclear matter and the global one-body
dynamics.  The central $\xenon+\tin$ collisions at 50 MeV/nucleon are
calculated by the models with and without shrinking, and it is shown
that the inclusion of the wave packet shrinking has a large effect on
the multifragmentation in a big expanding system with a moderate
expansion velocity.
\end{abstract}

\maketitle

\section{Introduction}
In order to describe heavy ion reactions as the dynamics of
many-nucleon systems, two different kinds of microscopic approaches
have been proposed and applied.  One is the molecular dynamics models
\cite{AICHELIN,MARUb,ONOab,FELDMEIER} and the other is the mean field
(TDHF-like) models \cite{BERTSCH,CASSING}.  An advantage of the mean
field models is that they do not put any restriction on the one-body
motion, while their disadvantage is that they cannot properly describe
the cluster formation because of the lack of many-body cluster
correlations.  On the other hand, usual molecular dynamics models
assume a fixed Gaussian shape for single particle wave functions.
This is an efficient way to describe the cluster correlation even by
using a simple product wave function with or without
antisymmetrization.  However, in another sense, the use of localized
wave packets can be a regression because the one-body dynamics is not
as precisely described as in mean field models.

An unified understanding is desired on the question whether the single
particle wave functions should be unrestricted or localized.  Unless
we can solve the dynamics keeping the full order of the many-body
correlations, it is essential for a reasonable model to introduce the
fluctuations which bring the system into many quantum branches each of
which corresponds to one of the reaction channels or the
configurations of clusterization.  In mean field models, it has been
proposed to introduce fluctuations in the one-body distribution
function \cite{AYIK,RANDRUP}, though it should be a difficult problem
to determine the correlations among an infinite number of the degrees
of freedom of fluctuations.  In this viewpoint, the philosophy of
molecular dynamics is to introduce a special kind of fluctuation by
stochastically localizing the single particle wave functions, which is
essential for the cluster production.  The mean field equation should
be interpreted as giving the short-time evolution of the one-body
distribution averaged over the stochastic branches.  Based on this
idea, the antisymmetrized molecular dynamics (AMD) has been extended
in Refs.\ \cite{ONOh,ONOi} by incorporating the wave packet diffusion
effect in the mean field as a source of the fluctuation to the wave
packet centroid which causes the quantum branching on the many-body
level.  Thus the single particle wave functions are localized in each
branch, which makes cluster formation possible, while the single
particle motion is not restricted for the averaged value over the
branches.  The stochastic treatment of the dynamics of the wave packet
width is the essential point of our approach.  There is another
approach \cite{FELDMEIER,KIDERLEN,CHOMAZ,MARU-EQMD} which treats the
width parameters as time-dependent variables in molecular dynamics,
though it has turned out that the deterministic dynamics of the width
variables cannot explain the evolution of the density fluctuation and
the multiple cluster formation \cite{KIDERLEN,CHOMAZ}.  Ohnishi and
Randrup \cite{OHNISHI-RANDRUPb,OHNISHI-RANDRUPc} have introduced
quantum fluctuation into wave packet molecular dynamics based on
their statistical consideration and have shown its importance in
cluster formation.  However the dynamical origin of the their
fluctuation has not been made clear.

An unsatisfactory point of the improvement in Refs.\ \cite{ONOh,ONOi}
was that the stochastic fluctuation to the wave packet centroids can
diffuse the distribution but cannot shrink the distribution.  Note
that a wave packet in the mean field normally diffuses in three
directions in phase space and shrinks in the other three directions.
Because of this difficulty, for example, the improved AMD could not be
directly applied to an isolated nucleon, and therefore the diffusion
was switched off for isolated nucleons, which introduces an ambiguity
to the model.  To solve this kind of problems, we absolutely need a
consistent understanding of both the mean field description and the
molecular dynamics description.

The first purpose of the present work is to construct a general
framework which contains both molecular dynamics models and mean field
models as specific cases.  In this framework, the time evolution of a
many-body system is given by the coherent mean field propagation and
the decoherence of single particle states into wave packets.  It will
have two physically essential ingredients, $\tau$ and
$\tau_{\text{mf}}$, which define when ``decoherence'' and ``mean field
branching'' take place, respectively.  The choice
$(\tau,\tau_{\text{mf}})=(\infty,\infty)$ corresponds to a mean field
model, while the choice $(\tau,\tau_{\text{mf}})=(0,0)$ corresponds to
the version of AMD of Refs.\ \cite{ONOh,ONOi} (called AMD/D in this
paper).  Based on this general framework, we introduce a new model
AMD/DS as the case of $(\tau,\tau_{\text{mf}})=(\text{large},0)$, with
which we can respect not only the diffusion but also the shrinking of
the phase space distribution predicted by the coherent mean field
propagation.

The second purpose of the present work is to demonstrate the effect of
the wave packet shrinking.  The difference between AMD/D and AMD/DS is
expected to result in the different diffusion properties of nucleons
in nuclear matter and the different global one-body dynamics.  We
perform the AMD/D and AMD/DS calculations for the central $\xenon +
\tin$ collisions at 50 MeV/nucleon and compare the results.  The
velocity of the expansion strongly depends on the model.  The
different expansion velocity results in the different cluster size
distribution.  It is shown, by comparison with INDRA experimental
data, that AMD/D had problems of the overestimation of $Z=4,5,6$
clusters and the underestimation of $Z\gtrsim15$ clusters and these
problems are solved by AMD/DS.

This paper is organized as follows.  In Sec.\ \ref{sec:formulation},
the formulation is presented.  The physical principle of the general
framework is given in \ref{subsec:Principle}, and then the concrete
formulae are given in \ref{subsec:Formulation} which includes all the
details.  In \ref{subsec:SpecificModels}, we introduce specific models
such as AMD/D and AMD/DS, and give simple examples to show how our
formulation works for AMD/D and AMD/DS.  In Sec.\ \ref{sec:calc}, the
results of the calculations with AMD/D and AMD/DS are compared to each
other and to the INDRA data for central $\xenon+\tin$ collisions at 50
MeV/nucleon, so as to demonstrate the important effect of the wave
packet shrinking in AMD/DS in multifragmentation.  Section
\ref{sec:summary} is devoted to a summary.

\section{\label{sec:formulation}
Mean field propagation followed by decoherence}

\subsection{\label{subsec:Principle}
Principle}

First we give a general framework which includes both mean field
models and molecular dynamics models as limiting cases of model
parameters.  Originally we have reached this framework in the course
of an extension of AMD by the incorporation of good points of mean
field models.  However, the framework is given here from a more
general point of view as an approximation of quantum many-body
dynamics.

In what follows, the two-nucleon collision effect is not shown
explicitly for the brevity of presentation, but it is always
considered in all the practical calculations as a stochastic process
\cite{ONOab,ONOd}.

In this model, we will use (without specifying yet how to use it) the
AMD wave function which is a Slater determinant of Gaussian wave
packets \cite{ONOab},
\begin{equation}
\langle\mathbf{r}_1\ldots\mathbf{r}_A|\Phi(Z)\rangle=
\det_{ij}\Bigl[\exp\Bigl\{
  -\nu\Bigl(\mathbf{r}_j - \frac{\mathbf{Z}_i}{\sqrt\nu}\Bigr)^2
\Bigr\}
\chi_{\alpha_i}(j)\Bigr].
\label{eq:AMDWaveFunction}
\end{equation}
The complex variables $Z\equiv\{\mathbf{Z}_i;\ i=1,\ldots,A\}$
represent the centroids of the wave packets.  We take the width
parameter $\nu=0.16$ $\mathrm{fm}^{-2}$ and the spin isospin states
$\chi_{\alpha_i}=p\uparrow$, $p\downarrow$, $n\uparrow$, or
$n\downarrow$.  Because of the antisymmetrization, the variables $Z$
do not necessarily have a direct physical meaning as nucleon positions
and momenta.  The AMD wave function $|\Phi(Z)\rangle$ contains many
quantum features in it and has been applied to the study of nuclear
structure problems with some extensions such as the parity and angular
momentum projections \cite{ENYO}.

The time-dependent many body wave function $|\Psi(t)\rangle$
describing a complicated nuclear collision is, in an intermediate or
final state, a superposition of a huge number of channels each of
which corresponds to a different clusterization configuration.  We do
not try to directly treat such a complicated many-body state
$|\Psi(t)\rangle$ nor we do not approximate $|\Psi(t)\rangle$ by a
single AMD wave function $|\Phi(Z)\rangle$.  We rather approximate the
many-body density matrix $|\Psi(t)\rangle\langle\Psi(t)|$ by an
ensemble of many AMD wave functions,
\begin{equation}
|\Psi(t)\rangle\langle\Psi(t)|\approx
\int \frac{|\Phi(Z)\rangle\langle\Phi(Z)|}{\langle\Phi(Z)|\Phi(Z)\rangle}
     w(Z,t)dZ.
\label{eq:AMDEnsemble}
\end{equation}
A pure many-body state is approximated by a mixed state.  This
approximation itself will not be a serious drawback practically
because the nuclear collision dynamics is so complicated that one
cannot observe full many-body correlations to distinguish a pure state
from a mixed state.  The benefit of this approximation is that the
right hand side of Eq.\ (\ref{eq:AMDEnsemble}) still contains
nontrivial many-body correlations required in multiple cluster
formation, even though an ensemble of AMD wave functions is
sufficiently simple to be tractable numerically.  We should, of
course, define a reasonable time evolution of the weight $w(Z,t)$ or,
alternatively, stochastic trajectories of the variables $Z(t)$.

What is the physical idea when we use AMD wave functions
$|\Phi(Z)\rangle$ in Eq.\ (\ref{eq:AMDEnsemble})?  When multiple
cluster formation takes place, $|\Psi(t)\rangle$ is composed of a huge
number of branches.  The decomposition into branches should be done so
that, in each branch, the one-body distribution of each nucleon is
localized in one of the clusters.  Namely, a nucleon should not belong
to many clusters at the same time, to avoid non-integer mass numbers
of clusters.  [The width of wave packets $\nu$ has been chosen in such
a way.]  In turn, if nucleons are localized in each branch, the
clusters are naturally bound due to the mean field among localized
nucleons.  This idea exists behind the molecular dynamics models which
restrict each single particle wave function in each branch to a wave
packet.

The time evolution in our model is determined by two factors, the mean
field propagation and the decomposition into branches.  At a time
$t_0$, let us take one of the branches
$|\Phi(Z)\rangle\langle\Phi(Z)|$ from Eq.\ (\ref{eq:AMDEnsemble}).
This is justified because the time evolution of a branch is
independent of the others due to linear quantum mechanics.  We
consider the mean field propagation from $t_0$ to $t_0+\tau$,
\begin{equation}
|\Phi(Z)\rangle\langle\Phi(Z)|\rightarrow
|\Psi(\tau,Z)\rangle\langle\Psi(\tau,Z)|,
\label{eq:MFieldProp}
\end{equation}
where $|\Psi(\tau,Z)\rangle$ is the solution of the mean field
equation
\begin{equation}
i\hbar\frac{d}{dt}|\Psi(t,Z)\rangle=
H^{\mathrm{HF}}(t)|\Psi(t,Z)\rangle
\label{eq:TDHFEquation}
\end{equation}
with the initial condition $|\Psi(0,Z)\rangle=|\Phi(Z)\rangle$.  The
mean field Hamiltonian has a form
\begin{equation}
H^{\mathrm{HF}}(t)=
\sum_{i=1}^{A}h_i(t)=
\sum_{i=1}^{A}\biggr(\frac{\mathbf{p}_i^2}{2M}
+U(\mathbf{r}_i,\mathbf{p}_i;\hat{\rho}(t))\biggr),
\end{equation}
where the potential $U$ depends on a one-body density matrix
$\hat{\rho}(t)$.  For the moment, we may assume that $\hat{\rho}(t)$
is the one-body density matrix for the state $|\Psi(t,Z)\rangle$.  We
wish to emphasize that the single particle wave functions are not
restricted to Gaussian packets in the mean field propagation and
therefore $|\Psi(\tau,Z)\rangle$ is a general Slater determinant.
Next, at the time $t_0+\tau$, the propagated state
$|\Psi(\tau,Z)\rangle\langle\Psi(\tau,Z)|$ is decomposed into AMD wave
functions as
\begin{equation}
\frac{|\Psi(\tau,Z)\rangle\langle\Psi(\tau,Z)|}
     {\langle\Phi(Z)|\Phi(Z)\rangle}
\approx
\int \frac{|\Phi(z)\rangle\langle\Phi(z)|}{\langle\Phi(z)|\Phi(z)\rangle}
      w(z,\tau;Z)dz,
\label{eq:Decomp2Branches}
\end{equation}
with a suitable weight $w(z,\tau;Z)$.  A reasonable principle to
determine $w(z,\tau;Z)$ would be first to choose a set of important
(one-body) operators $\{\hat{O}_\alpha\}$ and then to require that both
sides of Eq.\ (\ref{eq:Decomp2Branches}) give the same expectation
values for all $\{\hat{O}_\alpha\}$,
\begin{equation}
\int \frac{\langle\Phi(z)|\hat{O}_\alpha|\Phi(z)\rangle}
          {\langle\Phi(z)|\Phi(z)\rangle}
 w(z,\tau;Z)dz
=\frac{\langle\Psi(\tau,Z)|\hat{O}_\alpha|\Psi(\tau,Z)\rangle}
      {\langle\Phi(Z)|\Phi(Z)\rangle}.
\label{eq:WeightRequirement}
\end{equation}
However, this prescription will not always work well because it is
incompatible to the necessary condition $w(z,\tau;Z)\ge0$.  A practical
choice of $\{\hat{O}_\alpha\}$ will be given bellow.  In this way, in
principle, Eq.\ (\ref{eq:MFieldProp}) and Eq.\
(\ref{eq:Decomp2Branches}) determine the time evolution from $t_0$ to
$t_0+\tau$.  The next time evolution after $t_0+\tau$ is obtained
successively by applying the same model to each term of the right hand
side of Eq.\ (\ref{eq:Decomp2Branches}).

What is the physical meaning of $\tau$?  Is Eq.\
(\ref{eq:Decomp2Branches}) just a numerical approximation of a mean
field model or does it have any physical meaning?  Once one believes
that the mean field propagation were always perfect, then the choice
of a finite $\tau$ would be unphysical.  But the mean field
propagation is not perfect at least in that it unphysically keeps the
idempotency $\hat{\rho}^2=\hat{\rho}$ of the one-body density matrix.
In many-body systems, even though the idempotency is satisfied at a
time, it is not the case at a later time due to many-body
correlations.  The reduced one-body density matrix will be an ensemble
of idempotent density matrices.  If we ignore the antisymmetrization
for the simplicity of the discussion here, the one-body density matrix
of a nucleon will be given like
\begin{equation}
\hat{\rho}=w'|\varphi'\rangle\langle\varphi'|
+w''|\varphi''\rangle\langle\varphi''|+\cdots,
\label{eq:OneBodyDecoherence}
\end{equation}
in which different components $|\varphi'\rangle$,
$|\varphi''\rangle$,\ldots\ do not interfere in a simple way. The
many-body state for this situation will be like
\begin{equation}
|\Psi\rangle=c'|\varphi'\rangle\otimes|\tilde{\Psi}'\rangle
            +c''|\varphi''\rangle\otimes|\tilde{\Psi}''\rangle+\cdots,
\label{eq:OneBodyDecoherenceInManyBody}
\end{equation}
where the states of the rest of the system $|\tilde{\Psi}'\rangle$,
$|\tilde{\Psi}''\rangle$, \ldots\ are orthogonal to one another, to
have Eq.\ (\ref{eq:OneBodyDecoherence}).  In our model, the coherent
mean field propagation
($|\Psi\rangle=|\varphi\rangle\otimes|\tilde{\Psi}\rangle$) is assumed
to be valid in the time duration from $t_0$ to $t_0+\tau$, and
``decoherence'' is assumed to physically take place into Gaussian
packets at $t_0+\tau$ as in Eq.\ (\ref{eq:OneBodyDecoherence}) and
Eq.\ (\ref{eq:OneBodyDecoherenceInManyBody}).  [``Decoherence'' is a
general concept of quantum mechanics in open systems \cite{ZUREK}.  It
means not only that Eq.\ (\ref{eq:OneBodyDecoherence}) is satisfied at
a specific time but also that different branches do not interfere at
any later time.]  In this way, the parameter $\tau$ has a deep
physical meaning as the coherence time.  We do not try to determine
$\tau$ a priori in the present paper, but we assume that decoherence
is a physical process as should be the case at least in multiple
cluster formation.  It should be noted that the physical state changes
by decoherence, and therefore Eq.\ (\ref{eq:WeightRequirement}) should
not be required for all the operators $\{\hat{O}_\alpha\}$.  We should
require Eq.\ (\ref{eq:WeightRequirement}) for those operators
$\{\hat{O}_\alpha\}$ which we can believe to remain after physical
decoherence.

Closely related to the decoherence of a nucleon, we should decide how
this nucleon interacts with the rest of the system, namely, how this
nucleon contributes to the mean field potential $U$.  Once decoherence
takes place at $t_0+\tau$ so that different branches in Eq.\
(\ref{eq:OneBodyDecoherenceInManyBody}) do not interfere any longer,
then the mean field should be calculated in each branch, independently
of the other branches, by using the corresponding wave packet (one of
$|\varphi'\rangle$, $|\varphi''\rangle$, \ldots).  We will call this
change of the mean field ``mean field branching''.  The decomposition
of the many body state in Eq.\ (\ref{eq:Decomp2Branches}) is
consistent to the mean field branching at $t_0+\tau$.  Nevertheless,
we may think of the other possibility that the time scale of mean
field branching, denoted by $\tau_{\text{mf}}$, is shorter than
$\tau$.  The choice $\tau_{\text{mf}}<\tau$ can be reasonable in such
a physical situation that, even before decoherence, the mean field
approximation is applicable not to $|\Psi\rangle$ of Eq.\
(\ref{eq:OneBodyDecoherenceInManyBody}) but to each of the
``pre-branches'' in the right hand side of Eq.\
(\ref{eq:OneBodyDecoherenceInManyBody}) even when the pre-branches
have not been decohered yet (i.e., $|\tilde{\Psi}'\rangle$,
$|\tilde{\Psi}''\rangle$, \ldots\ are not orthogonal).  This situation
is possible because the true time evolution is linear while the mean
field approximation is nonlinear.  Therefore we regard
$\tau_{\text{mf}}$ as a physical ingredient of the model which is not
necessarily equal to $\tau$.

The physical origin of decoherence we have in mind here is the full
oder of many-body correlations beyond the two-body correlation, which
is especially important in multiple cluster formation.  The choice of
wave packets as decohered states is done for this purpose.  Of course,
the two-body collision effect destroys the idempotency of the one-body
density matrix, but this effect is already taken into account by a
more explicit way of the stochastic two-nucleon collision process
\cite{ONOab,ONOd} in all the realistic calculations.

\subsection{Formulation\label{subsec:Formulation}}

Now we start to give concrete formulae for the calculation of the time
evolution governed by the mean field propagation followed by the
decoherence into stochastic branches.  The explanation will be given
by five steps.

Impatient readers may be able to skip this part first moving to Sec.\
\ref{subsec:SpecificModels}, and later come back here if necessary.

\subsubsection{Physical coordinate approximation}
We adopt the ``physical coordinate approximation'' for an AMD wave
function,
\begin{equation}
\frac{|\Phi(Z)\rangle\langle\Phi(Z)|}{\langle\Phi(Z)|\Phi(Z)\rangle}\approx
\bigotimes_{k=1}^A \frac{|\varphi(\mathbf{W}_k)\chi_{\alpha_k}\rangle
                         \langle\varphi(\mathbf{W}_k)\chi_{\alpha_k}|}
                        {\langle\varphi(\mathbf{W}_k)\chi_{\alpha_k}|
                         \varphi(\mathbf{W}_k)\chi_{\alpha_k}\rangle}
\label{eq:PhysCoordApprox}
\end{equation}
where the spatial wave function of each nucleon $k$ is given by a
Gaussian packet
\begin{equation}
\langle\mathbf{r}|\varphi(\mathbf{W}_k)\rangle =
\exp\Bigl\{
  -\nu\Bigl(\mathbf{r} - \frac{\mathbf{W}_k}{\sqrt\nu}\Bigr)^2
\Bigr\}.
\end{equation}
The centroids are the physical coordinates $W=\{\mathbf{W}_k\}$
defined in Ref.\ \cite{ONOab} by
\begin{gather}
  \mathbf{W}_k
  =\sqrt{\nu}\,\mathbf{R}_k+\frac{i}{2\hbar\sqrt{\nu}}\mathbf{P}_k
  =\sum_{j=1}^A \Bigl(\sqrt Q\Bigr)_{kj}\mathbf{Z}_j,
  \label{eq:PhysCoord}\\
  Q_{kj} =B_{kj}B^{-1}_{jk},\qquad
  B_{kj}=e^{\mathbf{Z}_k^*\cdot\mathbf{Z}_j}\delta_{\alpha_k\alpha_j}.
\end{gather}
In the phase space representation, the Wigner function for the nucleon
$k$ is given by a Gaussian packet
\begin{equation}
g(x;X_k)=8\exp\biggr[-2\sum_{a=1}^6
 (x_a-X_{ka})^2\biggr],
\label{eq:PacketInPhaseSpace}
\end{equation}
where we have introduced the 6-dimensional phase space coordinates
\begin{gather}
x=\{x_a;\ a=1,\ldots,6\}=\biggl\{\sqrt\nu\,\mathbf{r},\
\frac{\mathbf{P}_k}{2\hbar\sqrt\nu}\biggr\},\\
X_k=\{X_{ka};\ a=1,\ldots,6\}=\biggl\{\sqrt\nu\,\mathbf{R}_k,\
\frac{\mathbf{P}_k}{2\hbar\sqrt\nu}\biggr\}.
\end{gather}

We wish to emphasize that much of the fermionic feature remains even
though we adopt the physical coordinate approximation [Eq.\
(\ref{eq:PhysCoordApprox})].  This is because the value of $W$ carries
the fermionic information.  For example, in the $6A$-dimensional space
of $W$, there exists Pauli forbidden region where $W$ cannot take
value for any choice of the original coordinate $Z$ \cite{ONOab}.
Nevertheless, it depends on the purpose whether this approximation
gives a good result.  For example, it gives only a poor approximation
for the evaluation of the Hamiltonian and the mean field potential.
For the calculation of such quantities, we use the exact
antisymmetrized wave function $|\Phi(Z)\rangle$ or use a better
approximation scheme \cite{ONOi}.

\subsubsection{Mean field propagation
\label{subsubsec:MeanFieldPropagation}}
The mean field propagation [Eq.\ (\ref{eq:MFieldProp})] from $t_0$ to
$t_0+\tau$ is calculated based on the Vlasov equation with the
``Gaussian-Gaussian approximation'' whose exact meaning is given
bellow.

We require that the Wigner function $\bar{f}_k(x,t)$ of the nucleon
$k$ should satisfy the Vlasov equation
\begin{equation}
\frac{\delta \bar{f}_k}{\delta t}=
-\frac{\partial h}{\partial\mathbf{p}}
  \cdot\frac{\partial \bar{f}_k}{\partial\mathbf{r}}
+\frac{\partial h}{\partial\mathbf{r}}
  \cdot\frac{\partial \bar{f}_k}{\partial\mathbf{p}},
\label{eq:VlasovEquation}
\end{equation}
at least approximately, for the time interval from $t_0$ to
$t_0+\tau$, with the initial condition $\bar{f}_k(x,0)=g(x;X_k)$.  A
special notation of the time derivative ($\delta/\delta t$) is adopted
for the mean field propagation in order to avoid a future possible
confusion.  Of course, it would be an easy numerical task to solve
Eq.\ (\ref{eq:VlasovEquation}) directly.  However, in order to make
the later decoherence process as simple as possible, we take another
method by writing the Wigner function as the mean of the stochastic
virtual phase space distributions of deformed Gaussian shape,
\begin{align}
\bar{f}_k(x,t)&=\overline{g(x;X_k(t),S_k(t))}\nonumber\\
&=\int g(x;X,S_k(t))w_k(X,t)\frac{d^6X}{\pi^3},
\label{eq:VirtualDecomp}
\end{align}
with
\begin{equation}
g(x;X,S)=\frac{1}{8\sqrt{\det{S}}}
\exp\biggr[-\frac{1}{2}\sum_{ab=1}^6
                      S^{-1}_{ab}(x_a-X_{a})(x_b-X_{b})\biggr].
\end{equation}
Thus we have the virtually stochastic variables $X_{ka}(t)$ and
$S_{kab}(t)$ which represent the centroid and the shape, respectively,
of the virtually stochastic distribution $g(x,X_k(t),S_k(t))$.  The
initial condition for them can be given by $X_{ka}(0)=X_{ka}$ and
$S_{kab}(0)=\frac{1}{4}\delta_{ab}$ at the initial time $t_0$ [Eq.\
(\ref{eq:PacketInPhaseSpace})].  The stochasticity of $S_{kab}(t)$ is
not shown explicitly in Eq.\ (\ref{eq:VirtualDecomp}) because we will
see later that its stochasticity is weak.

We shall now determine the virtually stochastic time evolution of
$X_{ka}(t)$ and $S_{kab}(t)$.  It is first noticed that the Vlasov
equation can be applied to each component of Eq.\
(\ref{eq:VirtualDecomp}) as long as Eq.\ (\ref{eq:VlasovEquation}) is
linear in $\bar{f}_k$.  The time evolution of $g_k(x,X_k(t),S_k(t))$
by the Vlasov equation
\begin{equation}
\frac{\delta g}{\delta t}=
-\frac{\partial h}{\partial\mathbf{p}}
  \cdot\frac{\partial g}{\partial\mathbf{r}}
+\frac{\partial h}{\partial\mathbf{r}}
  \cdot\frac{\partial g}{\partial\mathbf{p}},
\label{eq:VlasovEquation4g}
\end{equation}
is characterized by the time evolution of the first and the second
moments of the distribution
\begin{gather}
\frac{\delta}{\delta t}{X}_{ka}(t)=\frac{\delta}{\delta t}
\int x_a g(x,t)\frac{d^6x}{\pi^3},
\label{eq:VlasovForCentroind}\\
\frac{\delta}{\delta t}{S}_{kab}(t) =
\frac{\delta}{\delta t}\int \Bigl(x_a-X_{ka}(t)\Bigr)
                       \Bigl(x_b-X_{kb}(t)\Bigr) g(x,t) \frac{d^6x}{\pi^3},
\label{eq:VlasovForVariance}
\end{gather}
in which $(\delta/\delta t)g(x,t)$ is given by Eq.\
(\ref{eq:VlasovEquation4g}).  The shape of the distribution at
$t+\Delta t$ is characterized by $S_{kab}(t)+(\delta/\delta
t){S}_{kab}(t)\Delta t$.  This symmetric matrix is diagonalized by an
orthogonal matrix as
\begin{equation}
S_{kab}(t)+\frac{\delta}{\delta t}{S}_{kab}(t)\Delta t
=\sum_c \lambda_c O_{ac} O_{bc}.
\end{equation}
The distribution is generally diffusing in some directions and
shrinking in the other directions in the phase space.  We will extract
the component that is diffusing beyond the original width of the wave
packet [Eq.\ (\ref{eq:PacketInPhaseSpace})] by
\begin{equation}
\Bigl(\frac{\delta}{\delta t}S_{kab}(s)\Bigr)_+
=\lim_{\Delta t\rightarrow 0}\frac{1}{\Delta t}
\sum_c\max\Bigl(0,\ \lambda_c-{\textstyle\frac{1}{4}}\Bigr)O_{ac}O_{bc},
\label{eq:SDotDiffusion}
\end{equation}
which is now taken into account, not by changing the variable
$S_{kab}(t)$, but by giving a virtual Gaussian fluctuation $\Delta
X_{ka}(t)$ to the centroid $X_{ka}(t)$ satisfying
\begin{subequations}
\label{eq:FlctProperty}
\begin{gather}
\overline{\Delta X_{ka}(t)}=0,\\
\overline{\Delta X_{ka}(t)\Delta X_{kb}(t')}
=\Bigl(\frac{\delta}{\delta t}S_{kab}(t)\Bigr)_+\delta(t-t').
\label{eq:FlctVariance}
\end{gather}
\end{subequations}
The equation of motion for $X_{ka}(t)$ with virtual stochasticity may
be written as
\begin{equation}
\frac{d}{dt}X_{ka}(t)=\frac{\delta}{\delta t}X_{ka}(t)+\Delta X_{ka}(t).
\label{eq:EqOfMotionForX}
\end{equation}
The equation of motion for $S_{kab}(t)$ is
\begin{equation}
\frac{d}{dt}S_{kab}(t)=\frac{\delta}{\delta t}{S}_{kab}(t)
-\Bigl(\frac{\delta}{\delta t}{S}_{kab}(t)\Bigr)_+,
\label{eq:EqOfMotionForS}
\end{equation}
which does not contain the diffusing component beyond the original
width because its effect has been counted as the fluctuation to
$X_{kab}(t)$.

We emphasize again that the stochasticity has been, at this stage,
introduced only as a numerical method to solve the mean field
propagation [Eq.\ (\ref{eq:VlasovEquation})].  By taking the mean of
the ensemble of stochastic distributions $g(x,X_k(t),S_k(t))$ [Eq.\
(\ref{eq:VirtualDecomp})], our solution will reproduce the
deterministic solution of Eq.\ (\ref{eq:VlasovEquation}), up to the
Gaussian-Gaussian approximation.  This approximation means that we
have introduced some restriction on $\bar{f}_k(x,t)$ by considering
only the Gaussian fluctuation to the centroid of deformed Gaussian
distribution $g(x,X_k(t),S_k(t))$.  Nevertheless, $\bar{f}_k(x,t)$ is
not restricted to a Gaussian form because the stochastic centroid
$X_k(t)$ can move on its own way in each stochastic realization.

\subsubsection{Decoherence}
At the time $t_0+\tau$, we will now perform the decoherence into AMD
wave functions [Eq.\ (\ref{eq:Decomp2Branches})], by decomposing the
wave function of each nucleon $k$,
\begin{equation}
\frac{|\psi_k(\tau)\rangle\langle\psi_k(\tau)|}
     {\langle\psi_k(\tau)|\psi_k(\tau)\rangle}
\approx
\int \frac{|\varphi(\mathbf{w})\rangle\langle\varphi(\mathbf{w})|}
          {\langle\varphi(\mathbf{w})|\varphi(\mathbf{w})\rangle}
      w'_k(\mathbf{w},\tau)d\mathbf{w},
\label{eq:Decomp2BranchesW}
\end{equation}
where $|\psi_k(\tau)\rangle$ stands for the state after the mean field
propagation from $t_0$ to $t_0+\tau$ with the initial condition of the
Gaussian packet $|\varphi(\mathbf{W}_k)\rangle$ at $t_0$.  In the
phase space representation, an equivalent equation is written as
\begin{equation}
\bar{f}_k(x,\tau)\approx\int g(x;X)w_k'(X,\tau)\frac{d^6X}{\pi^3},
\label{eq:Decomp2BranchesX}
\end{equation}
with the wave packet $g(x;X)$ defined by Eq.\
(\ref{eq:PacketInPhaseSpace}).  We are going to show that the weight
$w_k'(X,\tau)$ in this equation is given by the weight $w_k(X,\tau)$
defined by Eq.\ (\ref{eq:VirtualDecomp}) under a reasonable choice of
the requirement.

Before deriving it, we should notice two features of the variable
$S_{kab}(\tau)$ which represents the shape of the stochastic Gaussian
distribution $g(x;X_k(\tau),S_k(\tau))$.  The first feature is that
the stochasticity of $S_{kab}(\tau)$ is weak because it is only
through the indirect influence of the stochasticity of $X_{ka}(t)$
[Eq.\ (\ref{eq:EqOfMotionForS}) and Eq.\ (\ref{eq:EqOfMotionForX})],
and therefore
\begin{equation}
S_{kab}(\tau)\approx S_{kab}'(\tau)
\qquad\mbox{if}\ |X_{ka}(\tau)-X'_{ka}(\tau)|\lesssim1,
\end{equation}
for different stochastic realizations $(X_{ka}(t),S_{kab}(t))$ and
$(X_{ka}'(t),S_{kab}'(t))$.  The second feature is that the
eigenvalues of $S_{kab}(\tau)$ ($\lambda_1\le\cdots\le\lambda_6$) are
bound by 0 and $\frac{1}{4}$ as is evident from the way of the
construction [Eq.\ (\ref{eq:EqOfMotionForS}) with Eq.\
(\ref{eq:SDotDiffusion})] and normally three of them are equal to
$\frac{1}{4}$,
\begin{equation}
0<\lambda_1\le\lambda_2\le\lambda_3
\le\lambda_4=\lambda_5=\lambda_6=\frac{1}{4}.
\end{equation}
[These two features are valid if the coherence time $\tau$ is not
very long or the mean field Hamiltonian is not very different from a
quadratic form (with arbitrary curvatures) in the interesting phase
space region, as we will see later in simple examples.]  For
convenience, we can change the variable from $x$ to $y$, the latter
being defined so as to diagonalize $S_{kab}(\tau)$,
\begin{equation}
y_b=\sum_{a}O_{ab}x_a,\qquad
Y_b=\sum_{a}O_{ab}X_a,
\end{equation}
where $O_{ab}$ is the matrix to diagonalize $S_{kab}(\tau)$,
\begin{equation}
S_{kab}(\tau)=\sum_c \lambda_c O_{ac} O_{bc}.
\end{equation}

We now require that, integrated out the shrinking directions, the
distribution in the diffusing directions $(y_4,y_5,y_6)$
\begin{equation}
\int\bar{f}_k(y,\tau)\frac{dy_1dy_2dy_3}{\pi^{3/2}}
\end{equation}
should be unchanged by decoherence.  From Eqs.\
(\ref{eq:VirtualDecomp}) and (\ref{eq:Decomp2BranchesX}), it is easily
seen that the requirement is satisfied by choosing $w_k'(Y,\tau)$ so
that
\begin{equation}
\int w_k'(Y,\tau)\frac{dY_1dY_2dY_3}{\pi^{3/2}}
=\int w_k(Y,\tau)\frac{dY_1dY_2dY_3}{\pi^{3/2}}.
\end{equation}
This condition is, of course, fulfilled by taking
$w'(Y,\tau)=w(Y,\tau)$.

What about for the shrinking directions $(y_1,y_2,y_3)$?  In typical
cases, the width of the Wigner function $\bar{f}_k(y,\tau)$ in the
shrinking directions is comparable to $\lambda_1,\lambda_2,\lambda_3$
and smaller than $\frac{1}{4}$.  [This is because the fluctuation has
been given only to the diffusing directions and therefore the weight
distribution $w(Y,\tau)$ is narrow in the shrinking directions, which
is the case if $\tau\lesssim2\pi/\omega$, where $\omega$ is the
oscillation frequency corresponding to the curvature of the mean field
Hamiltonian.]  Therefore, by using Gaussian packets of width
$\frac{1}{4}$ and positive weight $w_k'$ in Eq.\
(\ref{eq:Decomp2BranchesX}), it is impossible to reproduce the
shrinking component of $\bar{f}_k(y,\tau)$.  This is a physical
consequence of the decoherence into wave packets.  After decoherence
has taken place, the shrinking disappears due to the uncertainty
principle in each branch, which should be regarded as a physical
change of the state by decoherence.  Therefore we shall never require
that the distribution in the shrinking directions $(y_1,y_2,y_3)$
would be kept unchanged by decoherence.  Instead, we require that the
width of $\bar{f}_k(y,\tau)$ in these directions is kept as close as
possible to the minimum value $\frac{1}{4}$.  As mentioned above, the
weight distribution $w(Y,\tau)$ in the shrinking directions is usually
narrow, and therefore the choice $w_k'(Y,\tau)=w(Y,\tau)$ does not
increase the width of $\bar{f}_k(y,\tau)$ in the shrinking directions
much beyond $\frac{1}{4}$.

The derived numerical procedure for decoherence is quite simple.  The
virtually stochastic variable $X_{ka}(\tau)$, obtained by the mean
field propagation procedure, is now given the physical meaning as the
wave packet centroid.  Accordingly, the variable of the shape is
replaced as
\begin{equation}
S_{kab}(\tau)\rightarrow \frac{1}{4}\delta_{ab},
\end{equation}
and the calculation is continued to the next mean field propagation.

\subsubsection{Mean field branching}

Equation (\ref{eq:VlasovEquation4g}) for the mean field propagation of
the nucleon $k$ contains the mean field Hamiltonian $h$ which depends
on the state of the other nucleons.  If one follows the original idea
of the mean field propagation, $h$ should be calculated as
$h[\bar{f}(t)]$ by using the Wigner function $\bar{f}_l(x,t)$ given by
equations similar to Eq.\ (\ref{eq:VirtualDecomp}) for $l=1,\ldots,A$.
At the time $t_0+\tau$ when decoherence takes place for single
particle wave functions, the mean field Hamiltonian will also change
as
\begin{equation}
h[\bar{f}(\tau)]\rightarrow h[g(X(\tau))],
\label{eq:MFBranching0}
\end{equation}
the latter being calculated for the wave packets $g(x,X_l(\tau))$.  We
will call this change ``mean field branching''.

We may think of the other possibility that mean field branching does
not necessarily take place at the same time as the decoherence of
single particle wave functions.  Let us introduce the time scale
$\tau_{\text{mf}}$ of mean field branching which can be different from
$\tau$, and generalize Eq.\ (\ref{eq:MFBranching0}) to
\begin{equation}
h(t)=\left\{
\begin{array}{ll}
h[\bar{f}(t)]&\qquad\text{for}\quad 0\le t<\tau_{\text{mf}}\\
h[g(X(t))]&\qquad\text{for}\quad \tau_{\text{mf}}\le t<\tau
\end{array}
\right..
\end{equation}
In fact, in a more elaborate theory, the decoherence of wave functions
may take place gradually during the time interval from $t_0$ to
$t_0+\tau$.  Then it can be reasonable in our model with sudden
branching to take $\tau_{\text{mf}}$ shorter than $\tau$; for example,
$\tau_{\text{mf}}=\frac{1}{2}\tau$.  As mentioned in Sec.\
\ref{subsec:Principle}, another physical possibility of
$\tau_{\text{mf}}<\tau$ is found if one considers the decomposition
like Eq.\ (\ref{eq:OneBodyDecoherenceInManyBody}) before the
coherence time $\tau$.  Even though the pre-branches have not
decohered, the mean field approximation may be applicable not to the
total state $|\Psi\rangle$ but to each of the pre-branches
$|\varphi'\rangle\otimes|\tilde{\Psi}'\rangle$,
$|\varphi''\rangle\otimes|\tilde{\Psi}''\rangle$, \ldots\ separately.
This situation is possible because the true time evolution is linear
while the mean field approximation is nonlinear.  If this is the
physical case, we should take $\tau_{\text{mf}}<\tau$.  It should also
be mentioned that, if the coherence time $\tau$ is short compared to
the time scale of the diffusion $1/[(\delta/\delta t)S_l]_+$, the
result does not depend on the choice of $\tau_{\text{mf}}$
$(\le\tau)$.

The extreme case of $\tau_{\text{mf}}=0$ is convenient for the
numerical calculation with a code based on molecular dynamics.  In
this case, we only need to calculate the mean field Hamiltonian
$h[g(X(t))]$ without performing the mean averaging in Eq.\
(\ref{eq:VirtualDecomp}) which would be a hard numerical task.
Another merit is that $h[g(X(t))]$ can be replaced with a precise mean
field Hamiltonian $h[\Phi(Z(t))]$ which is obtained from the fully
antisymmetrized AMD wave function $|\Phi(Z(t))\rangle$ without
employing the physical coordinate approximation.

\subsubsection{Equation of motion and energy conservation}

The equation of motion (\ref{eq:EqOfMotionForX}) should be written in
the original coordinates $Z$.  Furthermore, a special consideration
is necessary so as to ensure the total energy conservation.  No change
has been made since Ref.\ \cite{ONOi} and these problems are not
directly related to the main aim of this paper.  Therefore the readers
who are not interested in them can skip this part and go to Sec.\
\ref{subsec:SpecificModels}.  The prescriptions which have been given
in Ref.\ \cite{ONOi} are briefly summarized bellow.

Before writing down the stochastic equation of motion for the wave
packet centroids $Z$, several comments are necessary.  The
deterministic part of the equation is derived by the time-dependent
variational principle for the AMD wave function rather than using the
deterministic part of Eq.\ (\ref{eq:EqOfMotionForX}), though these two
ways should be almost equivalent.  The fluctuation $\Delta X_{ka}$ is
for the physical coordinate $X_{ka}$ and therefore it is necessary to
convert it to the fluctuation for the original coordinate $Z$.  This
is done by introducing a stochastic one-body quantity $\mathcal{O}_k$
that generates the fluctuation $\Delta X_k$ in the form of the Poisson
brackets, as shown in Ref.\ \cite{ONOi}.  It has been implicitly
assumed that there is no correlation among the fluctuations of
different wave packets.  However, some minimal correlations should
exist so that the conservation laws are satisfied.

The equation of motion for the centroids is written as
\begin{equation}
\begin{split}
\frac{d}{dt}\mathbf{Z}_i=\{\mathbf{Z}_i,\mathcal{H}\}
+\sum_{k=1}^{A}\biggl[
&\Bigl\{\mathbf{Z}_i,\;\mathcal{O}_k
        +\sum_m\alpha_{km}\mathcal{P}_m\Bigr\}_{\mathrm{C}_k}
\\
&+\mu_k\Bigl(\mathbf{Z}_i,\;\mathcal{H}
        +\sum_m\beta_{km}\mathcal{Q}_m\Bigr)_{\mathrm{N}_k}
\biggr],
\end{split}
\label{eq:AMDVEqOfMotion}
\end{equation}
where the Poisson brackets $\{\mathcal{F},\mathcal{G}\}$ and the inner
product of the gradients $(\mathcal{F},\mathcal{G})$ are defined by
\begin{gather}
\{\mathcal{F},\mathcal{G}\}=
\frac{1}{i\hbar}\sum_{i\sigma,j\tau}\biggl(
  \frac{\partial\mathcal{F}}{\partial Z_{i\sigma}}
  C^{-1}_{i\sigma,j\tau}
  \frac{\partial\mathcal{G}}{\partial Z_{j\tau}^*}
-
  \frac{\partial\mathcal{G}}{\partial Z_{i\sigma}}
  C^{-1}_{i\sigma,j\tau}
  \frac{\partial\mathcal{F}}{\partial Z_{j\tau}^*}\biggr),
\label{eq:PoissonBracket}
\\
(\mathcal{F},\mathcal{G})=
\frac{1}{\hbar}\sum_{i\sigma,j\tau}\biggl(
  \frac{\partial\mathcal{F}}{\partial Z_{i\sigma}}
  C^{-1}_{i\sigma,j\tau}
  \frac{\partial\mathcal{G}}{\partial Z_{j\tau}^*}
+
  \frac{\partial\mathcal{G}}{\partial Z_{i\sigma}}
  C^{-1}_{i\sigma,j\tau}
  \frac{\partial\mathcal{F}}{\partial Z_{j\tau}^*}\biggr),\\
C_{i\sigma,j\tau}=
\frac{\partial^2}{\partial Z_{i\sigma}^*\partial Z_{j\tau}}
\log\langle\Phi(Z)|\Phi(Z)\rangle\qquad\sigma,\tau=x,y,z.
\end{gather}
The subscripts $\mathrm{C}_k$ and $\mathrm{N}_k$ attached to these
brackets indicates that the consideration is limited to the sets of
nucleons $\mathrm{C}_k$ and $\mathrm{N}_k$, respectively, where
$\mathrm{C}_k$ stands for the cluster that includes the nucleon $k$
and $\mathrm{N}_k$ stands for a neighborhood of the nucleon $k$.  The
explicit definition is given in Ref.\ \cite{ONOi}.

The first term of Eq.\ (\ref{eq:AMDVEqOfMotion}) has been derived
based on the time-dependent variational principle, with $\mathcal{H}$
given by
\begin{equation}
  \mathcal{H}(Z)=\frac{\langle\Phi(Z)|H|\Phi(Z)\rangle}
                    {\langle\Phi(Z)|\Phi(Z)\rangle}
          -\frac{3\hbar^2\nu}{2M}A+T_0(A-N_\mathrm{F}(Z)).
  \label{eq:AMDHamil}
\end{equation}
The quantum Hamiltonian $H$ includes an effective two-body interaction
such as the Gogny force \cite{GOGNY} which can be density dependent.
The spurious kinetic energies of the zero-point oscillation of the
center-of-mass of the isolated fragments and nucleons have been
subtracted in Eq.\ (\ref{eq:AMDHamil}) by introducing a continuous
number of fragments $N_\mathrm{F}(Z)$ {}\cite{ONOab,ONOd}.  Without
this subtraction, the Q-values for nucleon emissions and
fragmentations would not be reproduced.  The parameter $T_0$ is
$3\hbar^2\nu/2M$ in principle but treated as a free parameter for the
adjustment of the binding energies.

The first term in the square brackets of Eq.\
(\ref{eq:AMDVEqOfMotion}) is the fluctuation due to $\Delta X_k$
generated by the stochastic one-body quantity $\mathcal{O}_k$ as
mentioned above, with the correction for the conservation of the three
components of the center-of-mass coordinate and those of the total
momentum (denoted by $\{\mathcal{P}_m\}$).  The Lagrange multipliers
$\{\alpha_m\}$ should be determined by
\begin{equation}
\{\mathcal{P}_l,\mathcal{O}_k\}_{\mathrm{C}_k}
+\sum_m \{\mathcal{P}_l,\mathcal{P}_m\}_{\mathrm{C}_k}\alpha_{km}=0.
\label{eq:Alphakm}
\end{equation}

The second term in the square brackets of Eq.\
(\ref{eq:AMDVEqOfMotion}) is the dissipation term to achieve the
energy conservation.  Since the dissipation term should not violate
the other conservation laws, the center-of-mass coordinate, the total
momentum and the total orbital angular momentum (denoted by
$\{\mathcal{Q}_m\}$) are kept constant by determining the Lagrange
multipliers $\beta_{km}$ by
\begin{equation}
(\mathcal{Q}_l,\mathcal{H})_{\mathrm{N}_k}
+\sum_m(\mathcal{Q}_l,\mathcal{Q}_m)_{\mathrm{N}_k}\beta_{km}=0.
\end{equation}
The monopole and the quadrupole moments in the coordinate and momentum
spaces are also included in $\{\mathcal{Q}_m\}$ when $\mathrm{N}_k$ is
composed of more than 15 packets, because the global one-body
quantities should be well described without the dissipation term due
to the way how the fluctuation is derived.  The parameter $\mu_k$ is
then determined by
\begin{equation}
\mu_k=
-\frac{\{\mathcal{H},\;\mathcal{O}_k+\sum_m\alpha_{km}\mathcal{P}_m\}
      _{\mathrm{C}_k}}
      {(\mathcal{H},\;\mathcal{H}+\sum_m\beta_{km}\mathcal{Q}_m)
      _{\mathrm{N}_k}}
\label{eq:Muk}
\end{equation}
in order to conserve the total energy $\mathcal{H}$.  Finally, we need
to avoid the problem that the fluctuation is finite even near the
ground state and therefore the energy conservation is impossible.
This is done by introducing a reduction factor $\gamma_k$ in front of
the square brackets of Eq.\ (\ref{eq:AMDVEqOfMotion}) near the ground
state (see Ref.\ \cite{ONOi}).

\subsection{Specific models and simple examples
\label{subsec:SpecificModels}}
Our framework includes two essential parameters $\tau$ and
$\tau_{\text{mf}}$ which represent the time scales for the
decoherence into wave packets and the mean field branching,
respectively.  Each of the following models can be regarded as
corresponding to a specific choice of $(\tau,\tau_{\text{mf}})$ [with
an additional simplification in the case of the original AMD].  One of
the main aims here is to compare the last two models (AMD/D and
AMD/DS) by taking simple examples.

\subsubsection{Mean field models}
It is needless to mention that the choice of
$\tau=\tau_{\text{mf}}=\infty$ corresponds to the mean field theory
which solves the mean field equation [Eq.\ (\ref{eq:TDHFEquation})]
for an given initial Slater determinant toward the final state for
large $t$.  In fact, in this case, our model is equivalent to solving
the Vlasov equation
\begin{equation}
\frac{\partial\bar{f}}{\partial t}=
-\frac{\partial h[\bar{f}]}{\partial\mathbf{p}}
  \cdot\frac{\partial \bar{f}}{\partial\mathbf{r}}
+\frac{\partial h[\bar{f}]}{\partial\mathbf{r}}
  \cdot\frac{\partial \bar{f}}{\partial\mathbf{p}},
\label{eq:VlasovMF}
\end{equation}
with the Gaussian-Gaussian approximation introduced in Sec.\
\ref{subsubsec:MeanFieldPropagation}.  The collision term
\cite{BERTSCH,CASSING} is not explicitly shown here for the brevity of
presentation.

It is worthwhile to mention the similarity and the difference between
our scheme and the mean field transport theory with fluctuation
\cite{AYIK,RANDRUP}.  In the latter theory, a stochastic term $\Delta
f$ is added to the mean field equation.  However, from the view point
of this paper, we can equivalently interpret that the one-body Wigner
function $\bar{f}$ follows the usual mean field equation
(\ref{eq:VlasovMF}) and, at the same time, the decomposition is made
by
\begin{equation}
\bar{f}(\mathbf{r},\mathbf{p},t)=
\sum_{\Delta f}w(\Delta f,t)
                \Bigl[\bar{f}(\mathbf{r},\mathbf{p},t)
                +\Delta f(\mathbf{r},\mathbf{p})\Bigr]
\end{equation}
which defines a scheme of mean field branching.  Nevertheless, we
encounter a problem if we want to interpret each stochastic
realization $\bar{f}+\Delta f$ as a decohered state.  A special
implementation of the fluctuation $\Delta f$ is necessary, because the
randomness of $\Delta f$ does not generally ensure the idempotency of
$\bar{f}+\Delta f$ and the eventual localization of it in phase space.

\subsubsection{The original AMD}
In the original version of AMD \cite{ONOab}, the change of the wave
packet shape in the mean field propagation [Eq.\
(\ref{eq:MFieldProp})] has not been considered.  This corresponds to
replacing Eq.\ (\ref{eq:VlasovForVariance}) with
\begin{equation}
\frac{\delta}{\delta t}{S}_{kab}(t) = 0,
\end{equation}
to have a constant shape $S_{kab}=\frac{1}{4}\delta_{ab}$.  Then there
is no branching due to decoherence, and we have a deterministic
equation of motion
\begin{equation}
\frac{d}{dt}\mathbf{Z}_i=\{\mathbf{Z}_i,\mathcal{H}\}
\label{eq:AMDEqOfMotion}
\end{equation}
instead of Eq.\ (\ref{eq:AMDVEqOfMotion}).  It should be noted,
however, that the two-body collision effect has been incorporated as a
stochastic process \cite{ONOab,ONOd} in addition to Eq.\
(\ref{eq:AMDEqOfMotion}) already in this original version, as well as
in all the other versions of AMD.


\subsubsection{AMD/D\label{subsec:old}}
In Refs.\ \cite{ONOh,ONOi}, the wave packet diffusion effect by the
mean field propagation has been incorporated into AMD as a source of
the stochastic branching of the wave packets.  This version of AMD
corresponds to the case of the strongest decoherence
($\tau=\tau_{\text{mf}}\rightarrow0$) in the present general
framework.  It is straightforward to confirm that taking the limit
$\tau\rightarrow0$ exactly results in the formulation of Ref.\
\cite{ONOi}.

This version of AMD, which is also called AMD-V, is called AMD/D in
this paper, because the wave packet diffusion (D) effect in the mean
field propagation [Eq.\ (\ref{eq:TDHFEquation}) or
Eq. (\ref{eq:VlasovEquation})] is taken into account as branching
while the shrinking of wave packets in some phase space directions is
discarded.  It should be noted that the shrinking is a result of the
coherent mean field propagation which disappears in the case of the
zero coherence time $\tau=0$.

It is difficult to judge a priori whether the choice of $\tau=0$ is
reasonable or not.  The answer will depend on the type of reaction
which we want to describe.  AMD/D has been applied, with a reasonable
success, to various reaction systems in the medium energy region not
only for nuclear collisions \cite{ONOh,WADAa,WADAb,QINZHI} but also
for nucleon induced fragmentation reactions \cite{TOSAKA}.  However,
these successes do not mean that AMD/D is valid for all kinds of
reactions.

We can a priori expect that AMD/D will fail if decoherence is not a
physical case.  The most simple and clear example is the dynamics of a
single nucleon moving in a one-body external potential
$U(\mathbf{r})$, in which decoherence does not exist because the
nucleon is not interacting with anything.  In this case, the exact
time-evolution $\psi(\mathbf{r},t)$ is given by the time-dependent
Schr\"odinger equation with an initial condition
\begin{math}
\psi(\mathbf{r},t_0)\propto\exp[-\nu
(\mathbf{r}-\mathbf{Z}/\sqrt{\nu})^2].
\end{math}
It is known that Eq.\ (\ref{eq:VlasovEquation}) with
$h(\mathbf{r},\mathbf{p})=\mathbf{p}^2/2M+U(\mathbf{r})$ yields the
exact quantum-mechanical time evolution of the one-body Wigner
function $\bar{f}(\mathbf{r},\mathbf{p},t)$ when the potential has a
quadratic form, including the case of a free nucleon or a nucleon in a
harmonic oscillator potential with arbitrary curvatures.

\begin{figure}
\includegraphics[width=0.9\columnwidth]{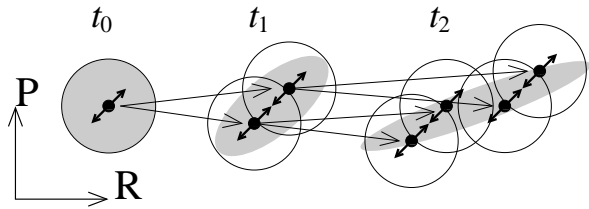}
\caption{\label{fig:WPNoShrink} The branching of the wave packet in
AMD/D is schematically shown for a free nucleon.  The
$\leftrightarrow$ symbols show the fluctuation to the wave packet
centroids.  Light gray region shows the exact time evolution of the
Wigner function $\bar{f}$.  }
\end{figure}

The light gray region in Fig.\ \ref{fig:WPNoShrink} shows the exact
time evolution of the Wigner function
$\bar{f}(\mathbf{r},\mathbf{p},t)$ for a free nucleon with the initial
condition of a Gaussian wave packet at $t_0$.  The spatial
distribution increases as the time progresses, while the momentum
distribution does not change at all.  Due to the Liouville theorem,
the phase space volume is conserved, which is the semiclassical
analogue of the fact that $\bar{f}(\mathbf{r},\mathbf{p},t)$
corresponds to a pure state $\psi(\mathbf{r},t)$.  At the initial time
$t_0$, the Wigner function is diffusing in the direction of the
$\leftrightarrow$ symbol in Fig.\ \ref{fig:WPNoShrink} and shrinking
in the other direction.  If AMD/D is applied to this situation, the
diffusion is taken into account as branching by giving fluctuation to
the wave packet centroid in the $\leftrightarrow$ direction, while the
shrinking is ignored.  At another time, $t_1$ or $t_2$, each of the
branched wave packets is treated in the same way as the initial wave
packet at $t_0$ (except for the different centroid value), without any
influence of the history of the wave packet and the existence of the
other branches.  Namely, the fluctuation always has the same property
for an free nucleon, and therefore both the spatial and the momentum
distributions increase as the time progresses.  Consequently the
coherent Wigner function $\bar{f}(\mathbf{r},\mathbf{p},t)$ cannot be
reproduced if AMD/D is applied.  Generally speaking, decoherence
increases the width of the phase space distribution.

In the practical calculations of nuclear reactions, the above problem
for a free nucleon is not so serious because the dynamics of free
nucleons is not of our interest, and it is usually sufficient to
describe an emitted nucleon as moving on a straight line (or on a
Coulomb trajectory) as a classical particle.  Therefore, in Refs.\
\cite{ONOh,ONOi} the branching was switched off for isolated nucleons.
For example, the switching-off condition adopted in Ref.\ \cite{ONOi}
is
\begin{subequations}
\label{eq:SwitchOffCondition}
\begin{equation}
\sum_i\theta\Bigl(1.75-|\Re(\mathbf{Z}_i-\mathbf{Z}_k)|\Bigr)\le10
\end{equation}
and
\begin{equation}
\biggl|\sum_i\theta\Bigl(1.75-|\Re(\mathbf{Z}_i-\mathbf{Z}_k)|\Bigr)
       \Re(\mathbf{Z}_i-\mathbf{Z}_k)\biggr|\le5.
\end{equation}
\end{subequations}
This condition switches off the branching for the nucleons in small
clusters with $A_{\text{c}}\lesssim10$ as well as emitted nucleons.

Not only the branching is switched off for isolated nucleons, but also
each isolated nucleon is regarded as having a definite momentum value
without the internal distribution of the Gaussian wave packet.  This
interpretation is consistent to the definition of the Hamiltonian
[Eq.\ (\ref{eq:AMDHamil})] where the zero-point oscillation kinetic
energies of isolated nucleons have been subtracted.  Therefore, we can
get reasonable results even with AMD/D if the switching-off condition
is appropriately chosen so that the branching is switched off (at
$t_2$ in Fig.\ \ref{fig:WPNoShrink} for example) when the momentum
centroid has got the appropriate amount of the fluctuation
corresponding to the internal momentum distribution of the initial
wave packet.

However, it will not be easy to find the switching-off condition that
works for all situations.  The condition of Eq.\
(\ref{eq:SwitchOffCondition}) may not work well for the reaction
systems which have not been studied.  When a big system is expanding
slowly, for example, the switching-off will not take place for a long
time, and then the branching will continue too long for the AMD/D
method to reproduce the mean field prediction.  In fact, we will show
in Sec.\ \ref{sec:calc} that the AMD/D method with the switching off
condition (\ref{eq:SwitchOffCondition}) seems to overestimate the
diffusion effect in a rather slowly expanding big system.  Therefore
we want such a new scheme of branching that we have no ambiguity of
introducing any switching-off condition.

\subsubsection{AMD/DS\label{subsec:new}}

Now we introduce a new scheme of decoherence by taking the choice of a
large coherence time $\tau$ and a short time scale for mean field
branching $\tau_{\text{mf}}=0$.  We will call this model AMD/DS
because the wave packet shrinking (S) effect in the mean field
propagation [Eq.\ (\ref{eq:TDHFEquation}) or
Eq. (\ref{eq:VlasovEquation})] is reflected in the dynamics as well as
the diffusion (D) effect with the choice of a finite $\tau$.  The
explicit definition of $\tau$, which depends on two-nucleon
collisions, will be given later.

\begin{figure}
\includegraphics[width=\columnwidth]{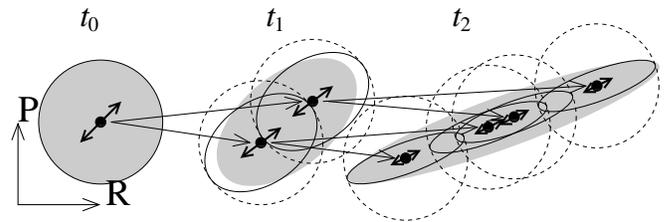}
\caption{\label{fig:WPWithShrink} The branching of the wave packet in
AMD/DS is schematically shown for a free nucleon.  Light gray region
shows the exact time evolution of the Wigner function $\bar{f}$.  Each
of the solid ellipses shows the shrunken shape
$g(\mathbf{r},\mathbf{p};X(t),S(t))$ in each branch, while the dashed
circles show the wave packets with the original width.  The
$\leftrightarrow$ symbols schematically show the magnitude and the
direction of the fluctuation to the wave packet centroids.}
\end{figure}

AMD/DS provides us a one-body dynamics similar to that in mean field
models, as expected from the fact that the coherent mean field
propagation is respected by the choice of a large coherence time
$\tau$.  Figure \ref{fig:WPWithShrink} illustrates how our formulation
in Sec.\ \ref{subsec:Formulation} works in the simplest example of a
free nucleon.  The light gray region is identical to that in Fig.\
\ref{fig:WPNoShrink}, showing the exact time evolution of the Wigner
function $\bar{f}(\mathbf{r},\mathbf{p},t)$ with the initial condition
of a Gaussian packet at the time $t_0$.  The wave packet diffusion
effect, $[(\delta/\delta t)S(t)]_+$ defined by Eq.\
(\ref{eq:SDotDiffusion}), is taken into account as before by giving
the fluctuation to the wave packet centroid
$X(t)=(\mathbf{R}(t),\mathbf{P}(t))$ in Eq.\ (\ref{eq:EqOfMotionForX})
with the property of the fluctuation given by Eq.\
(\ref{eq:FlctProperty}).  On the other hand, the shrinking effect is
reflected to the shape matrix $S(t)$ by the equation of motion
(\ref{eq:EqOfMotionForS}).  The ellipse with a solid line in each
branch in Fig.\ \ref{fig:WPWithShrink} represents the deformed and
shrunken shape of $g(\mathbf{r},\mathbf{p};X(t),S(t))$.  It should be
noted that the eigenvalues of $S(t)$ do not exceed the original wave
packet width $\frac{1}{4}$ (shown by dotted circles in Fig.\
\ref{fig:WPWithShrink}) since the diffusion beyond that width is
considered by the fluctuation to the centroid.  The exact Wigner
function $\bar{f}(\mathbf{r},\mathbf{p},t)$ is reproduced by the mean
average of the elliptic shapes $g(\mathbf{r},\mathbf{p};X(t),S(t))$ as
defined by Eq.\ (\ref{eq:VirtualDecomp}).

A general difference between AMD/DS and AMD/D is that the phase space
diffusion is weaker in AMD/DS than in AMD/D, reflecting the different
strength of decoherence.  Mathematically, this difference arises due
to the property of the fluctuation to the wave packet centroids.
Namely, the Vlasov equation (\ref{eq:VlasovEquation4g}) is applied to
different $g(\mathbf{r},\mathbf{p};X(t),S(t))$ which has always a full
width in AMD/D and has a shrunken shape in AMD/DS, and therefore the
property of the fluctuation [Eq.\ (\ref{eq:FlctProperty})] is
different.  In general, AMD/DS has smaller strength of the fluctuation
than AMD/D (except for the switching-off in AMD/D).  In the case of
the AMD/DS description of a free nucleon in Fig.\
\ref{fig:WPWithShrink}, as the time progresses, the strength of the
fluctuation gets smaller and smaller with the direction of the
fluctuation also changing, and at $t=\infty$ the fluctuation ceases
and the elliptic distribution $S(\infty)$ is completely shrunken to
have a definite value of the momentum.

It is easily proved that, in some cases, the coherent time evolution
of the Wigner function is reproduced exactly by AMD/DS as the mean
average of the shrunken elliptic shapes
$g(\mathbf{r},\mathbf{p};X(t),S(t))$, in spite of the
Gaussian-Gaussian approximation adopted in Sec.\
\ref{subsubsec:MeanFieldPropagation}.  This is the case not only for a
free nucleon but also for a nucleon in a harmonic oscillator potential
with arbitrary curvatures.  In general cases, however, the
reproduction is not exact because of the Gaussian-Gaussian
approximation.  In future, such an approximation may be removed if it
turns out to be necessary, but in the present work we are satisfied
with this agreement of AMD/DS, which ensures the approximate validity
of AMD/DS in the extreme case when decoherence does not physically
take place.

When the coherence time $\tau$ is long and the mean field branching
time is short ($\tau_{\text{mf}}=0$), it is interesting to imagine the
behavior of this model in nuclear collisions.  We can expect that the
mean field branching (namely, the stochastic fluctuation of the mean
field) does not influence so much the global dynamics and the early
dynamics in high density stage.  For such aspects, AMD/DS will behave
similarly to the mean field model because the coherence of the single
particle wave functions is kept for a large time scale $\tau$.  On the
other hand, for the aspect of clusterization, AMD/DS with
$\tau_{\text{mf}}=0$ will behave like usual molecular dynamics models,
because the branched mean field $h[\Phi(Z(t))]$ is equivalent to the
mean field in such models, and therefore it helps the formation of
clusters each of which is bound by the branched mean field.

The policy of the scheme of AMD/DS is to take $\tau$ that is as large
as possible.  However, if a two-nucleon collision takes place, it will
not make sense to keep the coherence of the single particle wave
functions of collided nucleons.  Therefore, we assume that decoherence
takes place for a nucleon with some probability when it experiences a
two-nucleon collision with another nucleon.  [Therefore $\tau$,
defined for each nucleon, is the time interval between two successive
collisions related to it.]  The probability $P_{\text{dec}}$ of
decoherence at each two-nucleon collision is chosen to be
\begin{equation}
P_{\text{dec}}(E,\theta)=e^{-E(1-\cos\theta)/E_0},
\label{eq:RestoreProb}
\end{equation}
where $E$ is the two-nucleon collision energy in the laboratory system
for the two nucleons and $\theta$ is the scattering angle in the
center-of-mass system for the two nucleons.  The purpose of this
probability is to reject the low momentum transfer cases where the
scattered state has a significant overlap probability with the case of
no collision.  Note that the probability is related to the momentum
transfer
\begin{math}
\mathbf{q}^2 
=ME(1-\cos\theta).
\label{eq:NNCollQsq}
\end{math}
The parameter $E_0=15$ MeV is chosen in the present work.  With this
choice, decoherence takes place in most of the collisions between the
nucleons from the different initial nuclei in the early stage of the
reaction with the incident energy more than several ten MeV/nucleon,
while decoherence seldom takes place within the initial nuclei and the
produced clusters.

At the end, let us discuss on the method of the subtraction of the
zero-point kinetic energies of the isolated nucleons in Eq.\
(\ref{eq:AMDHamil}).  This subtraction is consistent to the coherent
one-body dynamics of AMD/DS, in that the shrunken shape
$g(\mathbf{r},\mathbf{p};X(t),S(t))$ for an emitted nucleon eventually
has a definite momentum and therefore the zero-point kinetic energy
should not be counted in the conserved energy $\mathcal{H}$.  However,
when a nucleon is coming out of a nucleus, the zero-point energy
subtraction plays as a repulsive force to the nucleon from the nucleus
in spite of the fact that such a repulsion does not exist in the
one-body dynamics in the mean field.  It is possible to remove the
repulsive effect by keeping the conserved energy $\mathcal{H}$ given
by Eq.\ (\ref{eq:AMDHamil}).  This is easily done by formally adding a
term to the fluctuation part in the equation of motion
(\ref{eq:AMDVEqOfMotion}) by the replacement
\begin{equation}
\frac{\partial\mathcal{O}_k}{\partial Z_{j\tau}^*}
\rightarrow
\frac{\partial\mathcal{O}_k}{\partial Z_{j\tau}^*}
+
\delta_{jk}T_0\frac{\partial N_\mathrm{F}}{\partial Z_{j\tau}^*}.
\label{eq:NewSubZero}
\end{equation}
Then the added term cancels the zero-point subtraction term in the
deterministic part $\{\mathbf{Z}_i,\mathcal{H}\}$ and thus the
repulsive force mentioned above does not exist any longer.  The
replacement of Eq.\ (\ref{eq:NewSubZero}) is done at any place where
$\mathcal{O}_k$ appears, like in Eqs.\ (\ref{eq:Alphakm}) and
(\ref{eq:Muk}).  Therefore, the conserved energy is still
$\mathcal{H}$ with the zero-point energy subtracted.  The key of this
trick is that the zero-point energy is converted to the translational
motion of the nucleon in the old treatment while that energy is shared
by all the nucleons in the new treatment with the replacement of Eq.\
(\ref{eq:NewSubZero}).

\section{\label{sec:calc}
Effects in a Multifragmentation Reaction}

In this section, we discuss the results for $\xenon+\tin$ collisions
at the incident energy $E/A=50$ MeV and the impact parameter range $0
< b < 4$ fm.  A detailed and systematic analysis of this reaction
system will be given in separate papers
\cite{HUDAN-thesis,FRANKLAND-bormio}.  The main purpose here is to
make comparison of the two models of the quantum branching in AMD, one
(AMD/D) with only the wave packet diffusion effect and the other
(AMD/DS) with the wave packet diffusion and shrinking effects, the
latter effect being a consequence of the coherent mean field
propagation for a finite coherence time $\tau$.  From the character of
these models, as discussed in the previous section, we expect that
differences should be found in the diffusion property of nucleons in
nuclear matter and the global one-body dynamics.

Many events with various impact parameters in the range of $0<b<4$ fm
were produced by solving the stochastic equation of motion given in
Sec.\ \ref{subsec:Formulation}.  The triple loop approximation
\cite{ONOi} was used in order to save the computation time.  The Gogny
force \cite{GOGNY} was used as the effective interaction.  In the
calculation of AMD/DS, we use the new treatment of the subtraction of
the zero-point kinetic energies of nucleons and clusters given at the
end of Sec.\ \ref{subsec:new}, while AMD/D calculation is done with
the old treatment of the zero-point subtraction, unless otherwise
stated.  In addition to the equation of motion, the two-nucleon
collision effect was introduced in the usual stochastic way
\cite{ONOab,ONOd}.  The two-nucleon collision cross section adopted
here is given by
\begin{equation}
\sigma(E,\rho)=\min\biggl( \sigma_{\text{LM}}(E,\rho),\ \frac{100\
\text{mb}}{1+E/(200\ \text{MeV})}\biggr),
\end{equation}
where $\sigma_{\text{LM}}(E,\rho)$ is the cross section given by Li and
Machleidt \cite{LiMachleidt} from Dirac-Brueckner calculations using
the Bonn nucleon-nucleon potential.  This cross section depends on the
two-nucleon collision energy $E$ and the density around the collision
point $\rho$.  It also depends on the isospins of the colliding
nucleons.  The temperature in the parameterization by Li and Macheit
was, however, replaced by zero.  A low energy cut has been introduced
in the adopted cross section, though its effect on the final results
has turned out to be unimportant.  For the angular distribution, we
use the same parameterization as in Ref.\ \cite{ONOd}.

The calculation of each event was started by putting two nuclei with a
distance 12 fm and boosting them at the time $t=0$.  The AMD
calculation was continued until $t=300$ fm/$c$, at which we assume the
thermal equilibrium of each produced fragment and calculate its decay
by using a statistical decay code \cite{MARUb} which is based on the
sequential binary decay model by P\"uhlhofer {}\cite{PUHLHOFER}.

\begin{figure*}
\includegraphics[width=\textwidth]{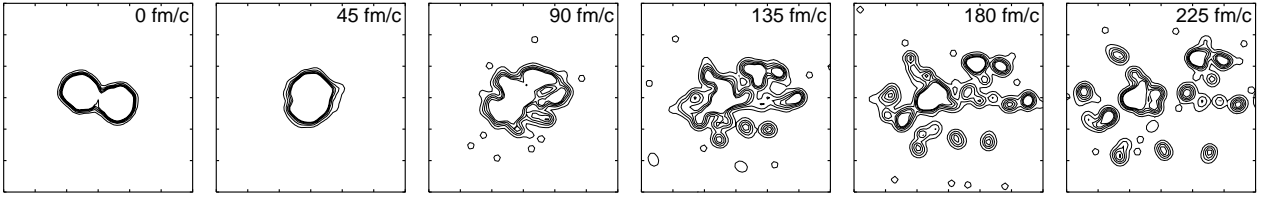}
\caption{\label{fig:anim} The time evolution of the density in the
center of mass system projected on to the reaction plane, in a typical
event of the $\xenon+\tin$ collision at 50 MeV/nucleon, from $t=0$ to
$t=225$ fm/$c$.  The beam direction is parallel to the horizontal
axis, and the impact parameter of this event is 3.4 fm. The size of
the shown area is $60\ \text{fm}\times 60\ \text{fm}$.  This is a
result of the calculation with AMD/DS.}
\end{figure*}

Figure \ref{fig:anim} shows the time evolution of a typical event with
the impact parameter $b=3.4$ fm.  It appears that a system is formed,
which, after a maximum compression around $t\sim45$ fm/$c$, expands
and many clusters appear around $t\sim 100$ fm/$c$.  The expansion is
stronger in the beam direction than in the transverse directions,
which means that the initial nuclei do not stop completely even in
such central collisions.  Therefore, another possible interpretation
may be that the initial two nuclei are passing each other with large
dissipation and breaking up into clusters.  However, the mixing of the
wave packets from the two nuclei is considerable.  On the average, 87
nucleons from the projectile $\xenon$ nucleus come out to the forward
direction ($p_z>0$ in the center-of-mass system), while the other 42
nucleons from the projectile nucleus appear in the backward direction.
This corresponds to around 67\%-33\% sharing of the projectile
nucleons in forward-backward directions, to be compared with 50\%-50\%
for full mixing or 100\%-0\% for no mixing at all.  These qualitative
features do not depend so much on the choice of the models of the
quantum branching, though the event of Fig.\ \ref{fig:anim} was
obtained with AMD/DS.

The same reaction system has been studied by Nebauer \textit{et al.}\
with the quantum molecular dynamics (QMD) \cite{INDRA}.  A serious
problem of their QMD result is that too large projectilelike and
targetlike fragments are produced even in the central collisions
($E_{\text{trans}}>450$ MeV).  Consequently the QMD calculation
largely overestimates the yield of the big clusters with $Z\gtrsim20$,
as shown in Fig.\ 7 of Ref.\ \cite{INDRA}.  This problem of the
spurious binary feature is qualitatively similar to the problem which
has ever encountered in the AMD calculation by Ono and Horiuchi for
the ${}^{40}\mathrm{Ca}+{}^{40}\mathrm{Ca}$ collisions at 35
MeV/nucleon \cite{ONOh}.  This problem in AMD has been solved in Ref.\
\cite{ONOh} by the stochastic incorporation the wave packet diffusion
effect which allows the mixing and/or breakup of the initial nuclei.
In fact, the present AMD calculation do not show the binary feature as
strong as in the QMD calculation, which can be seen in Fig.\
\ref{fig:anim} and in the cluster charge distribution to be shown
later.  The fermionic nature may also be important to solve the
problem of QMD.  Papa \textit{et al.}\ have shown in Ref.\ \cite{PAPA}
that the spurious binary feature disappears when they introduce the
fermionic nature into QMD in a stochastic way.

\begin{figure}
\includegraphics[width=\columnwidth]{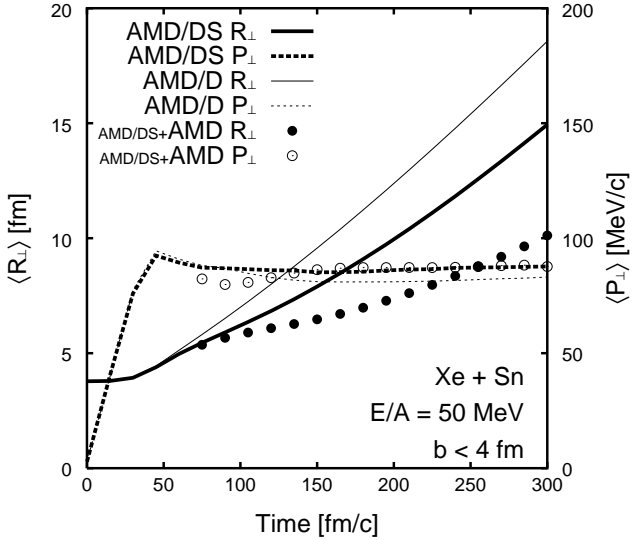}
\caption{\label{fig:expana} Time evolution of the transverse radius
$\langle R_\perp\rangle$ (solid lines) and the transverse momentum
$\langle P_\perp\rangle$ (dashed lines) in $\xenon+\tin$ collisions at
50 MeV/nucleon averaged for the impact parameter region $0<b<4$ fm.
The results of AMD/D are shown by thin lines, while those of AMD/DS
are shown by thick lines.  Filled and open circles show $\langle
R_\perp\rangle$ and $\langle P_\perp\rangle$, respectively, obtained
by the origial AMD calculation applied to the intermediate states at
$t=60$ fm/$c$ of the AMD/DS calculation.}
\end{figure}

We can expect that the early stage dynamics is not so sensitive to the
model of quantum branching, because AMD/D and AMD/DS are equivalent
for a short time scale and because the effective coherence time in
AMD/DS is short due to many two-nucleon collisions.  In fact, it is
seen in the part of $t\lesssim60$ fm/$c$ in Fig.\ \ref{fig:expana},
which shows the time dependence of the two quantities
\begin{align}
\langle R_\perp\rangle&=
\biggl\langle
\frac{1}{A}\sum_{i=1}^A \sqrt{R_{ix}^2+R_{iy}^2}
\biggr\rangle,\\
\langle P_\perp\rangle&=
\biggl\langle
\frac{1}{A}\sum_{i=1}^A \sqrt{P_{ix}^2+P_{iy}^2}
\biggr\rangle,
\end{align}
characterizing the transverse expansion.  We use the transverse
components of the physical positions $\mathbf{R}_i$ and the physical
momenta $\mathbf{P}_i$ defined by Eq.\ (\ref{eq:PhysCoord}).  The
brackets stand for the averaging over the events with the impact
parameter range $0<b<4$ fm.  [We consider these quantities rather than
the root mean square quantities in order to focus on the central part
of the system where clusters are mainly produced.]  The solid lines in
Fig.\ \ref{fig:expana} show the transverse radius $\langle
R_\perp\rangle$ and the dashed lines show the average transverse
momentum $\langle P_\perp\rangle$.  The thin lines show the result of
AMD/D, while the thick lines show the result of AMD/DS.  We can see
that the transverse momentum is produced in the early stage of the
collision (before 40 or 50 fm/$c$).  As expected, there is no
significant difference between AMD/D and AMD/DS for the early stage
dynamics $t\lesssim60$ fm/$c$.

Now our interest is in the evolution of the expanding system which has
been created by the early stage dynamics before $t\sim60$ fm/$c$.  In
Fig.\ \ref{fig:expana}, we can see that an important deviation between
the two models appears in the spatial radius $\langle R_\perp\rangle$
in later reaction stage.  The expansion velocity is slower in AMD/DS
with the wave packet shrinking effect than in AMD/D without it.  On
the other hand, the transverse momentum, which is almost constant for
$t\gtrsim60$ fm/$c$, is almost independent of the model of the quantum
branching.  It should be noted that the expansion is governed not only
by the momentum centroids of wave packets but also by the property of
the fluctuation to the wave packet centroids.  The latter is the
difference between the models, which can be naturally understood
because the wave packet shrinking effect reduces the strength of the
fluctuation to the wave packet centroids, as discussed in Sec.\
\ref{subsec:new}.  To see this effect more clearly, we also show, by
filled and open circles in Fig.\ \ref{fig:expana}, the results of the
original AMD without quantum branching.  In order to avoid the
influence of the different early stage dynamics, the original AMD was
applied to the intermediate states at $t=60$ fm/$c$ of the AMD/DS
calculation.  We can see that the transverse expansion is very weak
without quantum branching.

In order to get a deeper understanding, let us first consider how the
expansion dynamics is described if a mean field model (such as TDHF)
is applied.  If the two-nucleon collision effect is negligible in the
expanding system, most of the single particle wave functions will
widely spread over the space.  Clusters will not be produced in a mean
field model but the global one-body distribution may be reliable.  Due
to the coherence of the mean field propagation, the nucleon position
and the nucleon momentum are strongly correlated in the expanded
system, in a similar way to the case considered in Fig.\
\ref{fig:WPWithShrink} for example.  If we focus on a local part of
the expanded system, each nucleon has a rather sharp momentum
distribution like a classical particle.  The main aim of AMD/DS is to
have the same global one-body distribution as in mean field models,
when averaged over the branches.  The essential difference is that the
mean field varies from branch to branch in the case of AMD, which is
the reason why clusters are produced in AMD, though this difference
will not affect so much on the global one-body distribution.  When
clusters are formed, however, each nucleon is localized in one of the
clusters, and then it should have some momentum distribution that
satisfies the uncertainty relation and the Pauli principle.  If this
momentum distribution is considered, the global momentum distribution
will become wider than the mean field prediction.  Now the question is
whether this widening of the momentum distribution should be respected
or the coherent mean field propagation should be respected.  When we
apply AMD/DS, the coherent mean field propagation is respected and the
widening of the momentum distribution is not considered as far as the
one-body dynamics is concerned.  It may be possible that the wave
packet localization does not change the global one-body dynamics
through complicated many-body correlations which are out of the scope
of the present models.  On the other hand, when we take AMD/D, we
respect the widening of the momentum distribution due to the
localization of the wave packet as physical decoherence, which will
increase the future expansion velocity.  It is not possible to say a
priori that one model is superior to the other.  What we can say is
that AMD/DS reproduces the mean field prediction more precisely then
AMD/D.  It is another problem not discussed here whether the mean
field prediction is always reliable or not.

\begin{figure}
\includegraphics[width=\columnwidth]{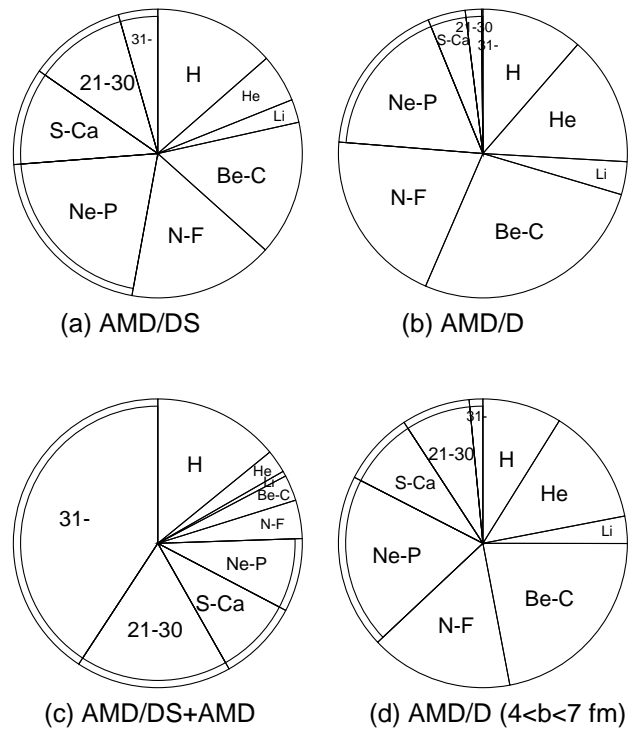}
\caption{\label{fig:zratio-dy} The partitioning of the total charge
into the clusters at $t=300$ fm/$c$ (before calculating the
statistical decay of excited clusters) in $\xenon+\tin$ collisions at
50 MeV/nucleon with the impact parameter $0<b<4$ fm.  The area of each
sector represents $ZM(Z)$ summed over the specified region of $Z$,
where $M(Z)$ is the multiplicity of clusters with charge $Z$.  (a) The
result of AMD/DS.  (b) The result of AMD/D.  (c) The result of the
original AMD applied to the intermediate states at 60 fm/$c$ of the
AMD/DS calculation.  (d) The result of AMD/D for the impact parameter
region $4<b<7$ fm.}
\end{figure}

\begin{figure}
\includegraphics[width=0.85\columnwidth]{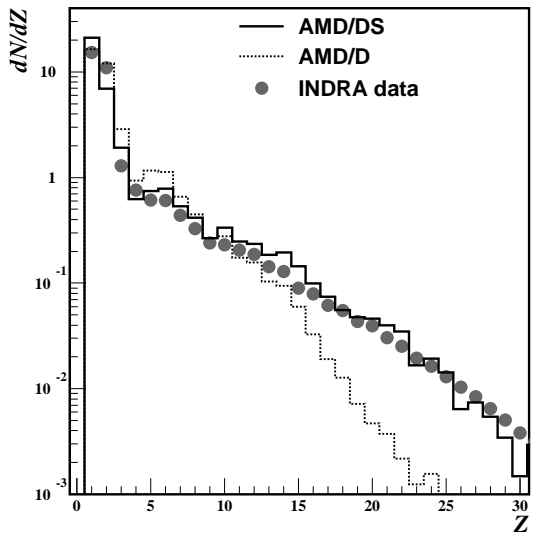}
\caption{\label{fig:zmulti} The charge distribution of the produced
clusters in $\xenon+\tin$ collisions at 50 MeV/nucleon with the impact
parameter $0<b<4$ fm, after calculating the statistical decay of
excited clusters and applying the experimental filter for the detector
setup.  Solid histogram shows the result of AMD/DS, while the dotted
histogram shows the result of AMD/D.  The INDRA experimental data
\cite{INDRA} are shown by bullets.}
\end{figure}

How does the different expansion velocity in the two models appear in
the observables?  As we have seen in Fig.\ \ref{fig:expana}, the
difference does not appear in the global transverse momentum, and
therefore the energy spectra are not good quantities to see the effect
directly.  The different expansion velocity is not due to the
different momentum but due to the different strength of the spatial
component of the fluctuation to the wave packet centroids.  Therefore,
we should look at the quantities which carry the information of the
increase rate of the spatial radius.  The cluster size distribution is
one of such quantities because each cluster is formed by the nucleons
with similar spatial positions and velocities.  It has been shown that
the cluster size decreases as the expansion velocity increases
\cite{GRADY,CHIKAZUMI}.  Figure \ref{fig:zratio-dy} shows how the
total 104 protons in the system are divided into clusters at $t=300$
fm/$c$ before calculating the statistical decay of excited clusters.
The results of AMD/DS and AMD/D are shown in (a) and (b),
respectively.  It is clearly seen that heavy clusters are produced
more abundantly in AMD/DS than in AMD/D, reflecting the different
expansion velocity.  In the case of the original AMD calculation
circles) linked to the early stage dynamics of AMD/DS [shown in (c)],
the produced clusters are much bigger than in AMD/DS, reflecting the
very slow expansion.  In Fig.\ \ref{fig:zmulti}, the final charge
distribution after statistical decay is shown together with the INDRA
data (bullets).  The data and the calculated results can be directly
compared, since the filter has been applied to the calculated events
in order to take account of the properties of the detector system.
The result of AMD/D has a serious problem that the multiplicity of the
heavy clusters with $Z\gtrsim15$ are underestimated.  Instead of the
heavy clusters, the relatively light clusters with $Z\sim 5,6$ are
produced too abundantly.  Therefore, it seems that the expansion is
too fast in AMD/D.  On the other hand, the reproduction by AMD/DS is
quite satisfactory for the charge distribution of the clusters with
$Z\ge3$.  These results suggest that it is reasonable to respect the
mean field prediction of the expansion dynamics in this reaction
system which consists of rather many nucleons and is expanding with a
moderate velocity.  However, AMD/DS has a problem of the
overestimation of the proton multiplicity and the underestimation the
$\alpha$ particle multiplicity.  Probably this is a side effect of the
fact that AMD/DS uses the mean field equation so faithfully that the
light cluster emission is not respected compared to the nucleon
emission.  A special care \cite{ONOcoal} will be necessary to explain
the direct production of light clusters which have only one quantum
bound state.  The use of the semiclassical version of the mean field
equation (the Vlasov equation) may also be one of the reasons why the
nucleon emission is overestimated.

The above quantitative results can be affected, in principle, by the
centrality selection method.  In our calculation all the events with
$0<b<4$ fm are considered as central events, while in experiment the
central events are selected by using the sum of the transverse
energies ($E_{\text{trans}}$) of observed light charged particles
($Z=1,2$).  The experimental data in Fig.\ \ref{fig:zmulti} were
obtained by selecting the events with the condition
$E_{\text{trans}}>450$ MeV \cite{INDRA}.  [These data are identical to
those which have been shown by the histogram in Fig.\ 7 of Ref.\
\cite{INDRA} in an arbitrary scale, while in our figure they are shown
in the absolute scale.]  Nebauer \textit{et al.}\ have shown in QMD
simulation that events with $4<b<6$ or 7 fm are also mixed in the
selected events with $E_{\text{trans}}>450$ MeV (Fig.\ 3 of Ref.\
\cite{INDRA}).  In order to estimate the effect of these
semi-peripheral events, we show in Fig.\ \ref{fig:zratio-dy}(d) the
charge partitioning obtained by AMD/D for $4<b<7$ fm.  We can see that
the character of clusterization in semi-peripheral events is not very
different from the central events [Fig.\ \ref{fig:zratio-dy}(b)].  The
difference between central and semi-peripheral events in AMD/D
calculation [(b) and (d)] is not as big as the difference between
AMD/D and AMD/DS in central events [(b) and (a)].  Therefore, a
possible considerable mixture of semi-peripheral events in the data
does not change our conclusion that too many small clusters are
produced in AMD/D and that the reproduction is improved by respecting
the coherence mean field propagation as in AMD/DS.

\begin{figure}
\includegraphics[width=0.85\columnwidth]{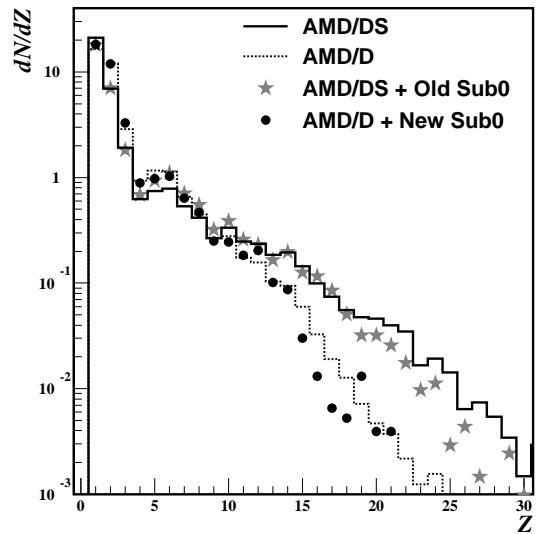}
\caption{\label{fig:zmulti-subzero} Similar to Fig.\ \ref{fig:zmulti}.
Solid and dotted histograms are the same as in Fig.\ \ref{fig:zmulti},
showing the results of AMD/DS and AMD/D, respectively.  The stars show
the result of the calculation with AMD/DS combined with the old
treatment of the zero-point kinetic energy subtraction.  The bullets
show the result of the calculation with AMD/D combined with the new
treatment of the zero-point kinetic energy subtraction.}
\end{figure}

Finally, the dependence on the treatment of the zero-point kinetic
energy subtraction is shown in Fig.\ \ref{fig:zmulti-subzero} for the
cluster charge distribution.  As far as the AMD/D model of the quantum
branching is used, the result is far from the experimental data
irrelevantly to the treatment of the zero-point subtraction.  In the
result of AMD/DS, we can get a slightly better reproduction of data by
using the new treatment of the zero-point subtraction given at the end
of Sec.\ \ref{subsec:new}.  It should be noted that the new treatment
of the zero-point subtraction is more consistent to the philosophy of
the AMD/DS model of the quantum branching in that the mean field
prediction of the one-body dynamics is respected.

\section{\label{sec:summary}
Summary}

In this paper, we have given a general framework which determines the
many-body quantum dynamics by the combination of the coherent mean
field propagation and the decoherence into branched wave packets.
This framework contains the mean field description and the molecular
dynamics description as specific cases.  The model given by Refs.\
\cite{ONOh,ONOi} (AMD/D) corresponds to taking the zero coherence time
($\tau=0$) for the mean field propagation.  In this scheme, the wave
packet diffusion by the mean field propagation is respected by giving
appropriate fluctuation to the wave packet centroids.  However, the
usual fluctuation was not able to describe the shrinking of the phase
space distribution which could be respected only by keeping the
coherence of the mean field propagation.  On the other hand, we have
shown in this paper that it is possible to implement a finite time
duration $\tau$ of the coherent mean field propagation before
decoherence, even though we still adopt a branching treatment.  As a
consequence, in the new model (AMD/DS), the shrinking of the phase
space distribution is respected as well as the diffusion.  AMD/DS
reproduces the exact dynamics for a free nucleon and for a nucleon in
a harmonic oscillator potential with arbitrary curvatures.  In general
cases, the branch-averaged one-body dynamics in AMD/DS should be much
closer to the prediction by mean field models than in AMD/D.
Nevertheless, by the choice of $\tau_{\text{mf}}=0$ for mean field
branching, clusters can be formed in AMD/DS as well as in usual
molecular dynamics models, because the mean field is calculated with
localized wave packets in each branch.  The two-nucleon collision
effect is introduced as usual, and the decoherence into wave packets
is assumed to take place when a nucleon experiences a two-nucleon
collision with a substantial momentum transfer.

The difference of the decoherence scheme between the two models
results in the different diffusion properties of nucleons in nuclear
matter and the different global one-body dynamics.  We have applied
both models of AMD/D and AMD/DS to the $\xenon+\tin$ collisions at 50
MeV/nucleon in the impact parameter range $0<b<4$ fm, where many
clusters are produced from the expanding system with a moderate
expansion velocity.  The effect of the wave packet shrinking in AMD/DS
certainly reduces the expansion velocity compared to AMD/D.
Reflecting this difference in the expansion velocity, the charge
distribution of the produced clusters strongly depends on the model of
decoherence into branches.  With AMD/DS, we have larger number of
heavy clusters with $Z\gtrsim15$ and smaller number of relatively
small clusters with $Z\sim5,6$ than with AMD/D.  AMD/DS reproduces the
INDRA experimental data much better than AMD/D, which suggests that
the coherent mean field propagation for the one-body dynamics should
be respected in this reaction system where a big system is expanding
with a moderate expansion velocity.  The detailed analysis of this
reaction system based on the AMD calculations will be given in a
separate paper.

However, we do not claim that AMD/DS is always superior to AMD/D or
vice versa.  These two models should be regarded as different schemes
of approximation.  AMD/DS respects the coherent mean field
propagation, while AMD/D respects the existence of strong many-body
correlations which causes the decoherence into branched wave packets.
Although the decoherence has been considered in AMD/DS based on the
two-nucleon collisions in this paper, it is also possible to have
other many-body effects to cause decoherence, with which AMD/DS can be
closer to AMD/D depending on the considered reaction systems.  In
future works, it will be important to investigate such possibilities.

\begin{acknowledgments}
 This work was supported by High Energy Accelerator Research
Organization (KEK) as the Supercomputer Projects No.\ 58 (FY2000) and
No.\ 70 (FY2001), and Le Commissariat \`a l'Energie Atomique, le
Centre National de la Recherche Scientifique and Le Centre de Calcul
du CEA, Grenoble under project number P542.  We also used the
supercomputer system at Research Center for Nuclear Physics (RCNP),
Osaka University.  A.~C.\ and A.~O.\ thank Patrick Bertrand for his
help installing the AMD software.
\end{acknowledgments}

\end{document}